\documentclass[12pt]{article}
\usepackage{graphicx}

\newcommand\pubnumber{CIPANP2018-Rozpedzik}

\def\napoli{Marian Smoluchowski Institute of Physics\\
Jagiellonian University, PL-31007, Cracow, Poland}
\def\leuven{Instituut
voor Kern-en Stralingsfysica, KU Leuven, B-3001, Leuven,
Belgium}
\def\ifj{Institute of Nuclear Physics, PAN, PL-31342, Cracow, Poland}
\def\Title#1{\begin{center} {\Large #1 } \end{center}}
\def\Author#1{\begin{center}{ \sc #1} \end{center}}
\def\Address#1{\begin{center}{ \it #1} \end{center}}

\newcommand\pubblock{\rightline{\begin{tabular}{l} \pubnumber\\
\\
\end{tabular}}}
\newenvironment{Abstract}{\begin{quotation}  }{\end{quotation}}
\newenvironment{Presented}{\begin{quotation} \begin{center} 
             PRESENTED AT\end{center}\bigskip 
      \begin{center}\begin{large}}{\end{large}\end{center} \end{quotation}}

\def\beq{\begin{equation}}
\def\eeq#1{\label{#1}\end{equation}}
\def\eeqn{\end{equation}}

\def\beqa{\begin{eqnarray}}
\def\eeqa#1{\label{#1}\end{eqnarray}}
\def\eeqan{\end{eqnarray}}

\let\bar=\overbar

\def\Dslash{\not{\hbox{\kern-4pt $D$}}}
\def\dslash{\not{\hbox{\kern-2pt $\del$}}}

\def\msb{{\bar{\ssstyle M \kern -1pt S}}}

\begin{document}
\begin{titlepage}
\pubblock

\vfill
\Title{Multi-Wire 3D Gas Tracker for Searching New Physics in Nuclear Beta Decay}
\vfill
\Author{D. Rozpedzik, K. Bodek, K. Lojek, M. Perkowski}
\Address{\napoli}
\Author{L. De Keukeleere, L. Hayen,  N. Severijns}
\Address{\leuven}
\Author{A. Kozela}
\Address{\ifj}
\vfill
\begin{Abstract}
Searches of new physics beyond the Standard Model (SM) performed at low energy frontiers are complementary to experiments carried 
out at high energy colliders. Among the methods for testing the SM and beyond at low energies are the precision spectrum shape and correlation 
coefficient measurements in nuclear and neutron beta decay. In order to study tiny effects in beta spectrum shape, a special spectrometer was built. 
It consists of a 3D low pressure gas tracker (drift chamber with hexagonal cells, signal readout at both wire ends) and plastic scintillators for 
triggering data acquisition and registration of the beta particle energy. The results of the characterization process indicate the possibility 
of using such a gas tracker in a range of experiments with low energy electrons where beta particle tracking with minimal kinematics deterioration is beneficial. 
Application of this technique is also planned for neutron decay correlation experiments. In the paper, the first application of this tracker in a 
high-precision beta spectrum shape study is discussed. The measurement technique, commissioning  results,  and the future outlook are
presented.
\end{Abstract}
\vfill
\begin{Presented}
CIPANP2018 - Thirteenth Conference on the Intersections of Particle and Nuclear Physics\\
Palm Springs, CA, USA, May 29--3 June, 2018 \\
\end{Presented}
\vfill
\end{titlepage}
\def\thefootnote{\fnsymbol{footnote}}
\setcounter{footnote}{0}

\section{Introduction}
In particle physics, the search for phenomena beyond the Standard Model (SM) is carried out on two energy frontiers.
High energy frontier is represented by complex experiments performed at high-energy accelerators. Low energy frontier is explored by rather small and high-precision
experiments. The results from these two domains can be compared thanks to the effective field theory. 
Searches for new effects beyond the SM at low energy include precision experiments with nuclear and neutron beta decay, in particular,
precision beta spectrum shape measurements and neutron correlation coefficients measurements.
In some of the $\beta$ decay correlation measurements there exist prospects to reach experimental
sensitivities better than $10^{-3}$ or even $10^{-4}$ making those observables interesting probes for searches of new physics 
originating at TeV scale. The most direct access to the exotic scalar and tensor interactions in nuclear and neutron $\beta$ decay is given by precision
measurements of the Fierz interference term (coefficient $b$). One way to access this term is through measurements of the shape of the  $\beta$
energy spectrum which should be performed with experimental accuracy below $10^{-3}$~\cite{hayen,nathal,nathal2,oscar}. The dominating contribution
 in the systematic uncertainty in such measurements comes from incomplete deposit of electron energy in detectors due to 
backscattering, transmission, bremsstrahlung effects.
Monte Carlo simulation of these effects is helpful, however, it introduces its own uncertainty as the input parameters are known with limited accuracy. 
The smallness of the potential $b$ contribution requires that other corrections to the spectrum shape of the same order are 
included in the analysis. While the $b$ coefficient contribution is inversely proportional to the total electron energy $E$, the recoil-order terms also affect the spectrum shape with their main contribution, from weak magnetism, being proportional to $E$. The weak magnetism is included in the SM but it is still poorly known. The difference in the energy dependence of the Fierz term and weak magnetism contributions in the $\beta$ spectrum shape can be exploited in order to distinguish these terms~\cite{hayen,oscar}. 

In this paper, we describe the present status of the multiwire 3D gas tracker which was build 
for studies of the experimental effects related to precision $\beta$ spectrum shape measurements. The gas tracker is the main part of the spectrometer called miniBETA.
Currently, many experiments try to measure the weak magnetism and Fierz terms.
A persistent problem in precision spectrum shape measurements are particle and energy losses in the detectors.
When a detector is hit by an electron its energy can be fully or partially deposited. If energy is partially deposited, it is crucial to identify and eliminate such events. 
To do this, application of a multiwire 3D gas detector for electron tracking can be helpful. 
Thanks to precise information about the electron track, it is possible to identify electrons backscatterred from the energy detector and eliminate electrons not originating in the $\beta$ source. Additionally, using the coincidence between a gas and an energy detector suppresses background from gamma emission typically accompanying beta decays. Light construction of the gas tracker helps reducing background from secondary radiation created inside the chamber due to collisions with the wires and mechanical support structure.
\section{Constraction of miniBETA spectrometer}

\begin{figure}[ht]
\centering
  \includegraphics[width=15cm,height=7cm]{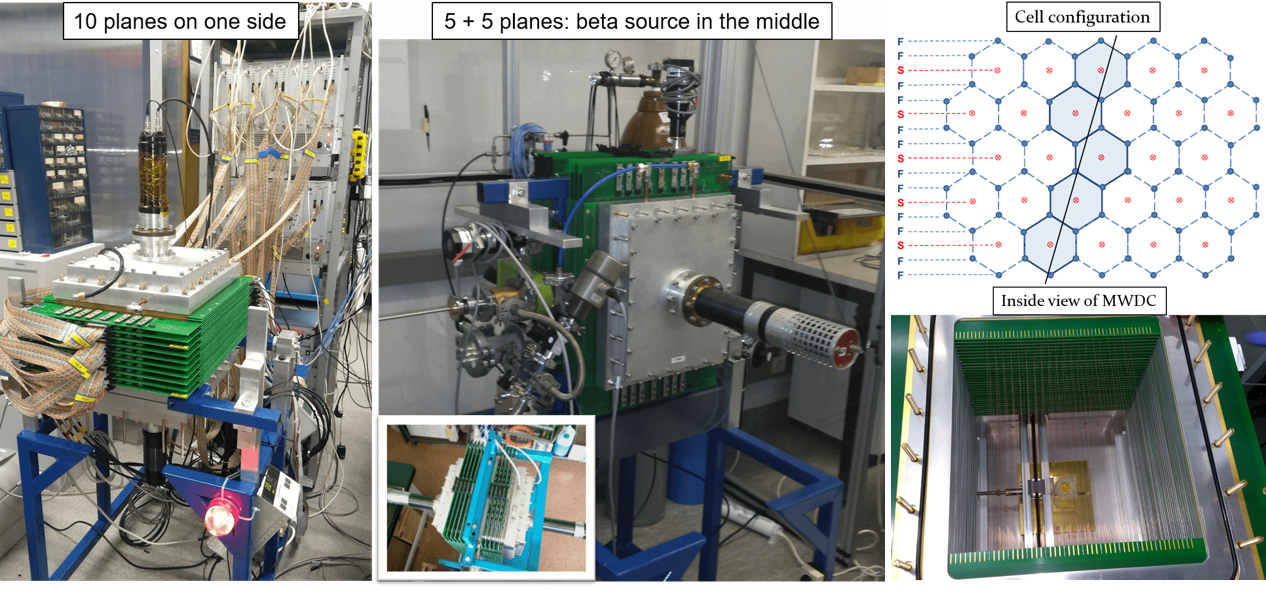} 
\caption{\label{fig-1}MiniBETA spectrometer. Leftmost panel: Horizontally oriented stack of 10 planes. Middle: vertically oriented source module mounted between two stacks of planes (5 +5). Rightmost panel: Geometry of wires in the hexagonal configuration. The blue (solid) circles denote the field wires and the red circles (crossed) correspond to signal wires. The field and signal wire planes are indicated with F and S, respectively. 
The shaded areas ilustrate the responding cells after an electron passed along the black solid line. }
\end{figure}

The main part of the miniBETA spectrometer is a low pressure multi-wire drift chamber which has a fully modular and reconfigurable design (see Fig.~\ref{fig-1}).

The reconfigurable design allows for two different cell geometries: rectangular and hexagonal. 
Both configurations try to satisfy the condition of maximum transparency of the detector and, consequently, a minimum number of necessary wires.
This condition is applied in order to minimize multiple electron scattering on the wires.  The MWDC design also allows to apply variable grouping of planes depending on the measurement purpose.
For studies of electron backscattering from the energy detector, a 5 + 5 plane configuration is preferred with a beta source in the middle of the chamber. 
Such a configuration allows for a good recognition of the so called "V--tracks" representing incoming and outgoing electron momenta. The present MWDC configuration, optimal for an overall performance characterization, consists of 10 sense wire planes mounted between 
24 field wire planes forming the double plane structure indicated in Fig.~\ref{fig-1}.
The distance between neighboring signal planes is 15 mm with wires within a plane separated by 17.32 mm. 
Each cell consists of a very thin anode wire (NiCr alloy, 25 $\mu$m diameter) surrounded by 6 cathode field wires (CoBe, 75 $\mu$m diameter) forming
a hexagonal cuboid. The anode wire materials are dictated by the charge division technique delivering information about the hit position coordinate along the wires.
All wires are soldered to pads of the printed circuit board (PCB) frames. The chamber is equipped with a source holder attached to the two--dimensional positioning system to be installed between suitable parts of the MWDC structure.  
Both end caps of the chamber are equipped with feedthroughs for lightguides connecting the in gas mixture embeded scintillator with a photomultiplier tube (PMT) installed outside the chamber. For the purpose of the MWDC characterisation, a plastic scintillator plate (Bicron BC400, 100x100x10 mm$^3$) was installed and read out by a single, $2"$ diameter photomultiplier tube. The miniBETA spectrometer was operated with He/isobutane gas mixtures ranging from 50/50 to 90/10 at pressures from 600 down to 300 mbar.

The MWDC is read out by a custom-designed modular electronic system. A single module consists of three main parts: preamplifiers, two analog cards containing the peak detector and threshold discriminator (THD), and the digital board containing analog-to-digital converters: ADC and TDC. One module serves one signal plane. The digital data is transmitted via a LAN port to the back-end computer where the process of receiving, sorting and formatting the data to a complete physical
event is accomplished by the data logging software. Plastic scintillators installed at both sides
of the MWDC provide the time reference signal for the drift time measurement.
The PMT signals are also used as a trigger for the readout sequence including MWDC data and information from electron energy detectors. 
A detailed description of the miniBETA DAQ can be found in Ref.~\cite{nim}. 
\section{3D track reconstruction algorithm}
\begin{figure}[h]
   \centering
   \includegraphics[width=11cm,height=5.5cm]{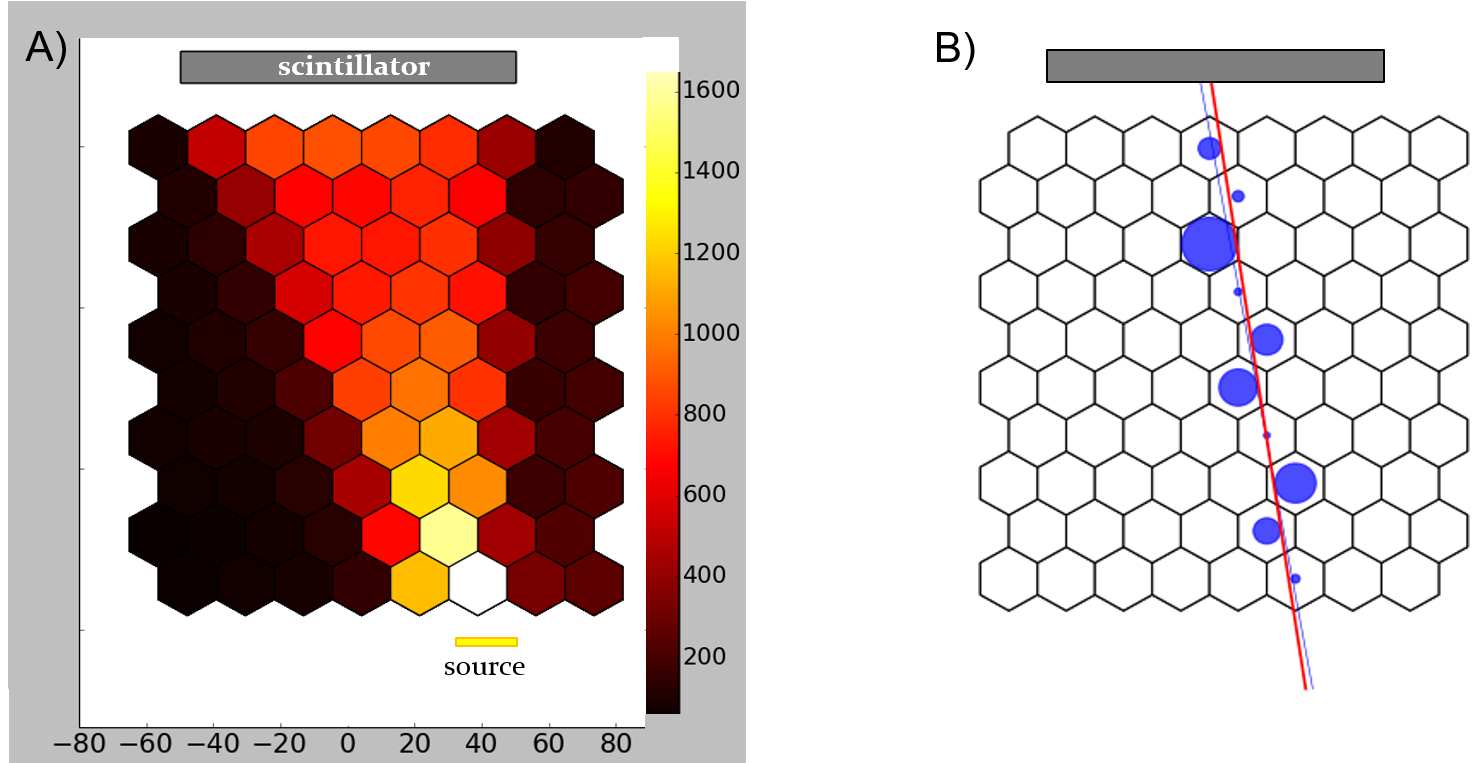} 
   \caption{\label{fig-2} (A) Illumination of MWDC cells by a beta source; (B) Example of the recognized and reconstructed electron track.
   The blue line represents the prefit and the red line represents the main fit. The circles originate in the centers of the cells which contributed to the event 
   and their radii $r_i$ are calculated according to the $r_i(t)$ calibration function described in Ref.~\cite{acta}.
}
\end{figure}

The data logging software~\cite{nim} provides the parallel reception of data frames sent by the electronics modules and the event building algorithm. 
The event structure includes full information about the event: unique time tag, ADC and TDC values from all wires as well as deposited energy in ADC units. 
Next, the analysis software reads the event structure and performs the reconstruction of electron tracks in three dimensions (3D). 
A dedicated algorithm was developed and implemented into the grafical interface for the data analysis. 
During measurements, this graphical interface allows for monitoring which cells fired and building intensity histograms as shown in Fig.~\ref{fig-2}A.
The 3D reconstruction algorithm consists of two 2D algorithms for reconstruction of the track projections in the XY and YZ planes. \pagebreak
Then the final 3D trajectory is defined by:
\begin{eqnarray}
x = ay + b \\
z = cy + d 
\end{eqnarray}
where $a$, $b$ are fit parameters from the XY projection plane based on the drift
time and $c$, $d$ are from the fit in the YZ plane based on the charge division method. 
For a given $y$ position one obtains a point $P_y (x,z)$ along the trajectory. 
Uncertainties of the track coordinates are calculated individually with the standard
error propagation procedure using the covariance matrices from fitting algorithms. 
The uncertainty of the $x$ coordinate is typically an order of magnitude lower than the one of $z$. 
An example of the reconstructed track in the XY plane is presented in Fig.~\ref{fig-2}B.  

The track reconstruction method based on the drift time information was described in detail in Ref.~\cite{acta}. 
Here we recall only the most important steps.
The XY reconstruction proceeds in following steps:
\begin{itemize}
\item{Cancellation of the "walk" effect is required due to the use of ordinary threshold discriminators for producing the STOP signal in the drift time measurement. The neccesary correction is obtained in a separate calibration procedure.}
\item{The relation between the drift time and the hit distance from the wire, for a given mixture, has to be found in advance, based on a dedicated measurement of cosmic muons.}
\item{In the prefit, a straight line through the centers of the cells that fired is fitted. This provides initial parameters for the main fit.}
\item{The main fit finds parameters of a straight line passing the closest to the circles originating in the selected cells centers with the radii based on their $r_i(t)$ calibration.} 
\item{A hit position ($x$,$y$) is found along the line for each of the cells that fired as the point with the smallest distance from the cell center. It is considered as the best approximation of the place where the primary ionization cluster giving the signal was produced. The $y$ coordinate of it is used for the YZ fit. }
\end{itemize}

In order to obtain a full 3D trajectory the $z$ coordinate of a hit in a cell has to be extracted from the place where the secondary ionization avalanche hits the anode wire.
The charge induced by the avalanche flows to both ends of the wire. The hit position can be estimated from the charge asymmetry collected on both wire ends.

The YZ reconstruction proceeds via the following steps:
\begin{itemize}
\item{Calculation and extraction of the ADC offsets $O_i$ (base-lines) for each channel.}
\item{Applying the gain corrections (from an independent calibration) in order to compensate uneven gains of the preamplifiers at both wire ends.}
\item{Application of the relation between the charge asymmetry and hit position obtained from an independent calibration. In general, this calibration function is nonlinear as shown in Ref.~\cite{nim}.}
\item{Fitting the electron trajectory in the YZ plane based on the linear regression to the collection of ($y$,$z$) points.}
\end{itemize}

\begin{figure}[h]
\centering
  \includegraphics[width=14.5cm,height=4cm]{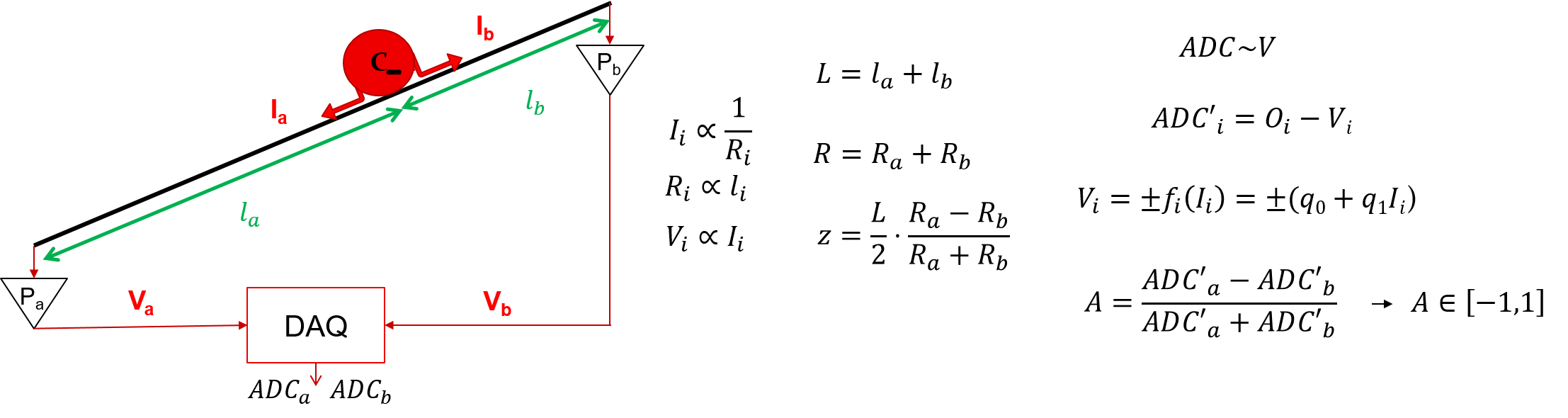} 
  \caption{\label{fig-4} Idea of the charge division method. A negative charge induced by an
avalanche flows to both ends of the wire generating a current $I_{a(b)}$ inversely proportional to the path length $l_{a(b)}$. At the end of each path a preamplifier $P_{a(b)}$ 
generates a differential voltage signal $\pm V_{a(b)}$, which is transmitted to the DAQ module and
converted into a digital signal.}
\end{figure}

The idea of the charge division technique is presented in Fig.~\ref{fig-4}.   
The current $I_i$ flowing in each path is inversely proportional to the path resistance $R_i$.
Wire resistivity and diameter are considered constant, hence the resistance is proportional
to the path length. 
Next, analog signals are transmitted to the DAQ module, where they are converted to digital signals and
recorded as $ADC_a$ and $ADC_b$ in ADC channel units. The incoming voltage signal of amplitude $V_i$ is measured in the DAQ module. 

\begin{figure}[!b]
   \includegraphics[width=15.2cm,height=3.6cm]{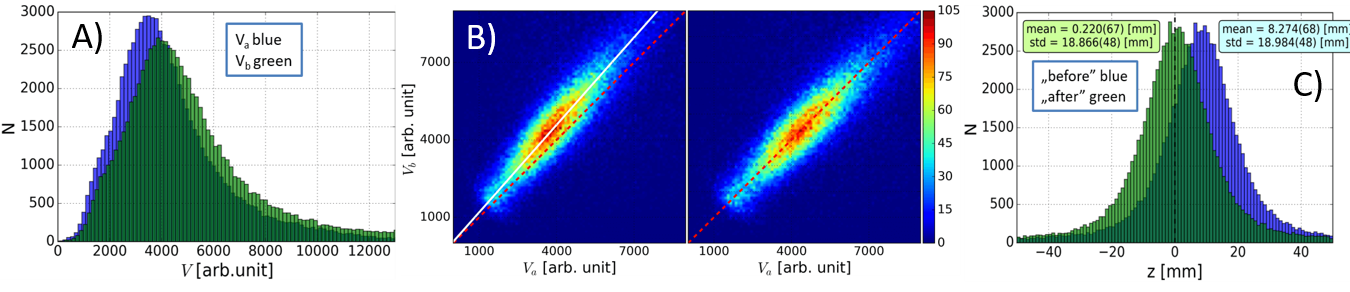}
   \caption{\label{fig-5} 
Results obtained from the gain correction dedicated measurements with the slit on the scintillator along z = 0 mm; 
(A) The measured $V_a$ (blue) and $V_b$ (green) distributions.
(B) On the left: A histogramed distribution of $(V_a, V_b)$ pairs.
The solid white line is the result of the ODR fit. The red dashed line represents the ideally balanced gains of $(V_a, V_b)$;
Corrected distribution is shown on the right;
(C) The distributions of calculated $z$ positions for a given wire
before (blue) and after (green) gain correction. }
\end{figure}

Preamplifiers at each end of the wire transform currents linearly into differential voltage signals $V_i$ depending
on the preamplier gain. Since the preamplifiers can have a different gain, one has to experimentally
determine the gain correction needed for each pair of preamplifiers used in
the chamber. Figure~\ref{fig-5}A presents measured distributions of values $V_a$ and $V_b$ 
for one wire before the correction. In panel B of Fig.~\ref{fig-5} the correlation histograms of 
$(V_a$, $V_b)$ are shown before and after correction.
If both preamplifiers had the same gain the distribution would be symmetric along
the line $V_a = V_b$. Fitting a straight line provides the two parameters ($p_{0}, p_{1}$) necessary for the gain correction 
for the given pair of preampliers. The line was fitted with the orthogonal distances regression (ODR) procedure,
which minimizes residua along both axes $(V_a$, $V_b)$.
The explicit equation for determining the $z_i$ position for the given wire $i$,
parameterized by experimentally found parameters is defined by:
\begin{equation}
z_i = \frac{L}{2} \frac{ADC'_{i,b} - (p_{i,0} + p_{i,1}ADC'_{i,a})}{ADC'_{i,b} + (p_{i,0}+ p_{i,1}ADC'_{i,a})}.
\label{eq1}
\end{equation}
In Fig.~\ref{fig-5}C the distribution of calculated $z$ values for a given wire
before (blue) and after (green) the gain ratio correction are compared. 
The measurement was performed with a 5 mm wide collimator centered in the middle of the wire. The calculated
gain corrections are then automatically applied for all further measurements.
This procedure has to be repeated every time preamplifiers are replaced in
the system.
\begin{figure}[!t]
   \centering
   \includegraphics[width=15cm,height=4cm]{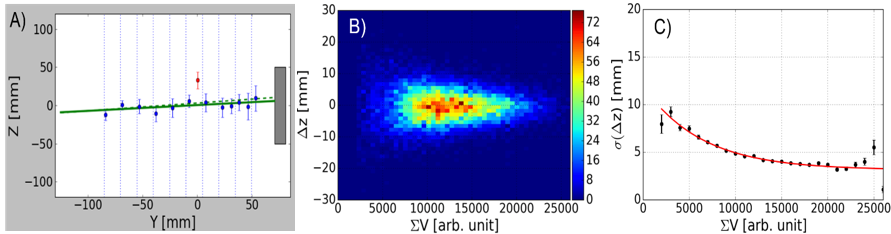} 
   \caption{\label{fig-6} (A) Fitted muon trajectory (green line) in YZ projection. 
Vertical dotted blue lines represent signal wires positions. The gray rectangle
on the right represents the scintillator. The dashed green line corresponds to
the initial fit including all hits (dots). Hit which was treated as outliers
is colored in red. A solid green line corresponds to the fit excluding outliers;
(B) Example of the 2D histogram of $(\sum V,\Delta z)$ pairs for the given
cell obtained for YZ trajectory fitting for cosmic muons;
(C) Example of the standard deviation $\sigma(\Delta z)$ of the ZY fit residua distribution dependency 
on the sum of recorded signals $\sum V$. The red line represents the fitted exponential function.
 }
\end{figure}

Figure~\ref{fig-6}A presents an example of a registered cosmic muon and its trajectory
fitted in YZ projection. Some outlier hits with $\Delta^2 z_i > 5~<\Delta^2 z_i>$ were removed from the fit procedure. 
It was observed that the sum of the signal amplitudes ($\sum V$) of recorded hits (proportional to the total charge)
has a major effect on their residua distributions ($\Delta z$). 
Figure~\ref{fig-6}B presents an example of a 2D histogram of $(\sum V,\Delta z$) pairs for the given cell. 
Position resolution along the $z$ direction was identified with the standard deviation of the distribution of the fit residuals. 
This distribution is shown in Fig.~\ref{fig-6}B while the $z$ resolution as a function of the sum of the signal amplitudes $(\sum V)$ is shown in Fig.~\ref{fig-6}C.
\section{Results of spectrometer characterisation}
\begin{figure}[!t]
\centering
   \includegraphics[width=15cm,height=9.3cm]{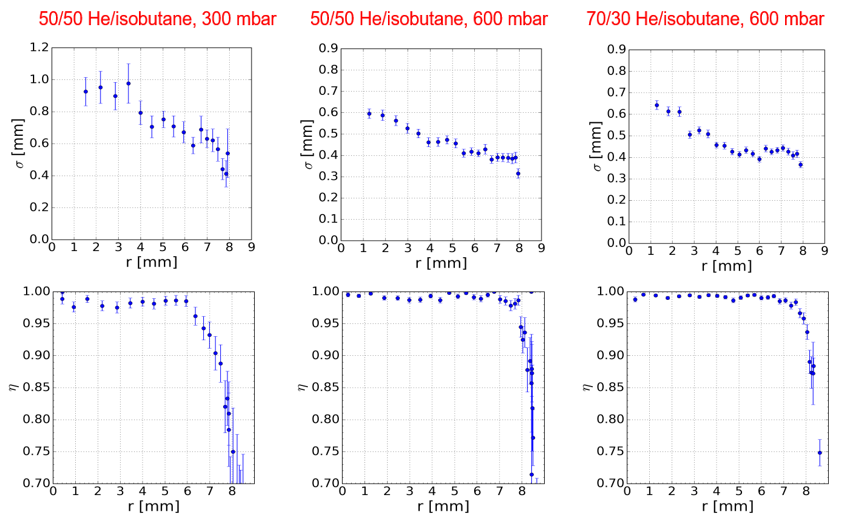}
\caption{\label{fig-7} Results obtained with selected He/isobutane gas mixtures ratios at fixed pressure. Upper row: Position resolution in the XY projection as a function of a distance $r$ 
from the anode wire.
Bottom row: The single cell efficiency as a function of the radius $r$. Uncertainties at the edge of the cell are higher due to a lower statistics.
}
\end{figure}

Figure~\ref{fig-7} shows part of the results obtained from the chamber characterization
using the different Helium/isobutane gas mixture ratios. Left column presents resolution and efficiency which was obtained with the mixture ratio set to 50/50 at 300 mbar pressure. 
Middle column includes the results with a gas mixture of 50/50 at 600 mbar pressure and right column is for 70/30 at 600 mbar pressure.
The drift chamber position resolution was
defined as the standard deviation $\sigma_{ij}$ of residua distribution for cosmic
muon trajectories using the final $r_i(t)$ calibration (see Ref.~\cite{acta}), where $i$ denotes the individual cell and $j$ relates to the selected subset of the drift times.
The resolution slightly declines for trajectories passing closer to the cell boundaries as the
cylindrical symmetry of the electric field breaks there. 
Moreover, the electric field is weaker closer to cells boundaries, hence electrons from the primary ionization
wander longer in these regions, can migrate to the neighboring cells or even
get lost. In addition, the field varies between cells, in particular for the outside cells. 
Therefore, the efficiency $\eta_i(r)$ was calculated individually for each cell as a function of the drift
radius. According to the procedure described in Ref.~\cite{acta} fitted trajectories of cosmic muons were used to calculate the chamber effciency.
In almost the entire cell $(r < 7$ mm) the single cell efficiency is constant at the level of 0.98 or better (see Fig.~\ref{fig-7}).
In addition, the total cells efficiencies was calculated for each cell. For most of the cells it was equal to 0.95 or greater.
It is important to note that the relatively low total efficiency is caused by the cell periphery being dominated
by a weak and nonsymmetric field.
\newpage
\section{Conclusions and outlook}
\begin{itemize}
\item{The miniBETA spectrometer consisting of a low-mass, low-Z materials 3D-tracker and a plastic scintillator for energy detection exhibits an average efficiency of the trajectory detection of about 98\% for most of the cells.}
\item{3D tracking performance -- position resolution: x and y $<$ 0.5 mm, z $<$ 6 mm.}
\item{Scattering effects in the chamber are inherently small due to: \\
-- 5 $\mu$m thick source foil\\
-- He/isobutane gas mixtures at the low pressure (100 mbar neo-pentane considered) \\
-- Hexagonal geometry with signal readout at both ends minimizes number of wires needed.}
\item{Further optimization of the plastic scintillator (light collection, gain uniformity) is needed – ongoing.}
\item{First physics goal: measurement of the weak magnetism term in $^{114}$In$\rightarrow^{114}$Sn and $^{32}$P$\rightarrow^{32}$S decays.}
\item{The miniBETA gas tracker can be usefull for all experiments where precision electron tracking is needed. In particular, measurement techniques used in miniBETA will be applied in the neutron decay correlation experiment – BRAND project~\cite{kaziu}.}
\end{itemize}


\begin{thebibliography}{99}

\bibitem{hayen}
L. Hayen et al., Rev. Mod. Phys. {\bf 90}, 015008 (2018).

\bibitem{nathal}
N. Severijns and  O.  Naviliat--Cuncic,  Phys.  Scr. {\bf T152}, 014018 (2013).

\bibitem{nathal2}
N. Severijns, J. Phys. G Nucl. Part. Phys. {\bf 41}, 114006 (2014).

\bibitem{oscar}
M. Gonzales--Alonso and O. Naviliat--Cuncic, Phys. Rev. C, {\bf 94},  035503 (2016).

\bibitem{nim}
K. Lojek et al. Nucl. Instrum. Meth. Phys. Res. A, {\bf 802}, 38 (2015).

\bibitem{acta}
M. Perkowski et al., Acta Physica Pol. A, {\bf 23}, 2647 (2018).

\bibitem{kaziu}
K. Bodek, Acta Phys. Pol. B, {\bf 47}, 349 (2016) and references therein.


\end{thebibliography}
\end{document}